%% file: main_submitted_un_normalized_dynamics.tex
\documentclass[journal]{new-aiaa} 
\usepackage[utf8]{inputenc}
\usepackage{graphicx}
\usepackage{amsmath,bm,bbm}
\usepackage{amsfonts} 
\usepackage[version=4]{mhchem}
\usepackage{siunitx}
\usepackage{multicol}
\usepackage{longtable,tabularx}
\usepackage{diagbox}
\usepackage{booktabs}
\usepackage{comment}
\usepackage{marvosym}
\usepackage{gensymb}

\graphicspath{{Figures/}} 
\input{command}

\title{Low-Thrust Orbital Differential Games \\ with Speed Constraint Enforcement Using Cost Weighting}
\author{Yahli Drucker\footnote{Graduate Student, The Stephen B. Klein Faculty of Aerospace Engineering, Technion, \href{mailto:yahlidrucker@campus.technion.ac.il}{yahlidrucker@campus.technion.ac.il}.} and  Vitaly Shaferman\footnote{Associate Professor, The Stephen B. Klein Faculty of Aerospace Engineering, Technion, \href{mailto:vitalysh@technion.ac.il}{vitalysh@technion.ac.il}.}}
\affil{Technion, Israel Institute of Technology, Haifa, Israel, 3200003}

\begin{document}
\maketitle

\section{Introduction}
\lettrine{O}{rbital} spacecraft engagements are complex and evolving challenges that have gained increasing attention in recent years. A reasonable approach to modeling such problems is to formulate a one-sided optimal control problem in which the controlled spacecraft chases a passive target in orbit while minimizing a predetermined cost function.
In \cite{lembeck1993optimal}, the one-sided approach was considered for the classic spacecraft rendezvous problem, i.e., reaching the target spacecraft with zero relative velocity. The problem was modeled using the Clohessy-Wiltshire (CW) equations, with the controller being the three-dimensional thrust-acceleration vector. A quadratic cost function, containing a running cost on the control effort (which represents the fuel consumption), was chosen, and optimal closed-loop guidance laws were derived for this problem, with either fixed or free final times. It was shown that the optimal final time, which minimizes the control effort, is unbounded, and a modified cost function that includes the final time was used to obtain optimal free-time solutions.
Another terminal constraint required for docking, for example, is a directional rendezvous (i.e., rendezvous from a prescribed direction). In \cite{guelman2001optimal}, this problem was similarly formulated as a two-stage process: first, the spacecraft was guided toward an intermediate point along the required approach direction, and then it was softly constrained to move straight along that direction. In \cite{nahum2024optimal}, the same problem was solved analytically by recasting it as a single optimization in which the directional constraint was replaced by passing through two intermediate points along the required approach direction.  

The main limitation of the one-sided approach is the assumption that the target is not maneuvering. When this assumption is violated, the guidance law's performance substantially decreases. A two-sided optimization problem, often known as a pursuit-evasion (PE) differential game, can be solved to overcome this limitation. The idea of modeling such engagements with a two-player differential game was first introduced by Isaacs in \cite{isaacs1965differential}. In this approach, the problem is formulated such that the pursuing spacecraft minimizes the cost function, while the evading spacecraft maximizes it. The optimal controller for each player is derived simultaneously, assuming the other player has perfect information and performs its optimal maneuver. The solution to both players’ optimal strategies is called a saddle point and is defined such that if either player deviates from its optimal strategy, it cannot gain.
In \cite{pontani2009numerical}, an orbital minimum-time, three-dimensional PE game was considered, based on the two-body problem's nonlinear equations of motion (EOMs). Saddle-point equilibrium solutions were obtained using a direct numerical optimization method, with an initial guess obtained via genetic algorithms.

Typically, direct methods handle complex problems more efficiently, converge more rapidly, and are less sensitive to the initial guess than indirect methods. However, they are usually less accurate than indirect methods that explicitly solve the necessary optimality conditions. Moreover, indirect methods can sometimes yield analytical solutions for simplified problems that direct methods cannot. 
In \cite{shen2018revisit}, the three-dimensional orbital PE problem was revisited. The same problem as in \cite{pontani2009numerical} was considered, but solved using an indirect method. The differential game problem was transformed into a two-point boundary value problem (TPBVP), which was solved using a classical shooting approach based on Newton’s method. 
In \cite{li2020saddle}, an indirect method was proposed for the problem of the three-dimensional orbital PE game under $J_{2}$ perturbed dynamics. The problem was formulated in terms of nonlinear relative orbital dynamics, including the two-body gravitational terms and the $J_{2}$ perturbation, with the terminal game time as the objective function. By combining shooting and collocation methods, the proposed approach enabled the robust identification of saddle points across various scenarios.
In \cite{innocenti2016game} and \cite{jagat2017nonlinear}, a nonlinear version of the relative motion equations and a quadratic cost function were used to derive a closed-loop optimal guidance law. The nonlinear equations were formulated in state-dependent matrix form, and the optimal controllers were obtained by solving the state-dependent Riccati equation. It was concluded that the solution may be near-optimal, as the state-dependent matrix form is non-unique.

The nonlinear nature of the two-body problem significantly complicates the solution of the differential games problem. However, the EOMs can be linearized by assuming that the distance between the spacecraft is small relative to the reference orbit's semimajor axis.
In \cite{pang2024solving}, the time-optimal PE game of two relatively close spacecraft near an elliptical reference orbit was considered. The problem was modeled using the Tschauner–Hempel equations \cite{tschauner1964optimale} and solved using either the damped Newton’s method or the homotopy method, depending on the maneuverability difference between the two spacecraft.

The problem can be further simplified by assuming that both spacecraft are in near-circular orbits, so that the CW equations apply. The CW equations constitute a linear, autonomous dynamical system that may admit analytic solutions via the indirect approach. Analytical solutions are particularly important for space applications, especially for small satellites, which often have limited computational resources. 
In \cite{stupik2012optimal}, a free-time constant-thrust differential game was considered for this case. The problem was formulated using the CW equations, with the final time as the cost function. Two numeric solutions for this problem were presented: an open-loop solution using particle swarm optimization, a heuristic, stochastic optimizer, and a closed-loop solution using kriging, a type of optimizer that uses pre-computed open-loop optimal solutions.
In \cite{dong2018satellite}, a fixed-time spacecraft interception guidance law using a differential games model in the CW frame was presented. The thrust magnitudes of both the pursuer and the evader were bounded, and the cost function was the relative terminal distance between the players. The zero-effort miss variables were defined via a coordinate transformation to reduce the order of the problem and simplify the solution. The derived solution had a bang-bang structure, typical of problems with only a terminal cost and bounded controls. Since the cost function in this problem did not include the control effort (or the final time, for a constant-thrust acceleration), the solution was not fuel-optimal.
In \cite{ye2020satellite}, a free-time guidance law for the PE differential game was presented. The game model was also based on the CW equations, but with the final time as the cost function. Each of the pursuer's thrust vector components was limited by a maximum thrust bound, whereas for the evader, the Euclidean norm of the thrust was bounded. It was concluded that for some initial conditions, indirect methods might not be able to solve this problem, unlike the situation where the Euclidean norm of the thrust is bounded for both players.

In all of the works mentioned thus far, the terminal velocity was either unconstrained or constrained to be zero. However, other velocity constraints, which are sometimes required, should also be considered.
For example, docking missions may require a slow, controlled approach at a specific low speed to activate the locking mechanism, while kinetic interception scenarios may require a high terminal speed.
In \cite{drucker2026optimal}, a one-sided linear-quadratic spacecraft engagement guidance law with terminal velocity constraints was derived. A zero-effort-miss coordinate transformation was used, and the terminal velocity was optimally selected to meet the user-defined terminal speed requirement.
To the best of our knowledge, this is the only optimal spacecraft-engagement guidance law that explicitly addresses non-zero terminal velocity constraints. Moreover, spacecraft-differential-game guidance laws with terminal velocity constraints other than zero have not been previously proposed.

This paper fills this gap in the literature. 
It derives a PE linear-quadratic differential games guidance law subject to a terminal speed constraint. It is assumed that both spacecraft are relatively close to each other and to a circular reference orbit.
The problem is posed with a quadratic cost function that includes a running cost on each player's control effort and terminal costs on the relative position and relative velocity.
The terminal speed is constrained by properly choosing the parameters of its weighting matrix, which, to the best of our knowledge, is a new approach. The solution for any terminal speed can be obtained by solving for the roots of a sixth-order polynomial. 
The weighting matrix associated with the terminal velocity cost can be either positive-definite (PD) or negative-definite (ND). The ND case that arises in the high-terminal-speed regime is rarely reported in the literature because it yields a nonconvex cost function.
A conjugate point analysis is presented to ensure the optimality of the solutions for both the PD and ND terminal velocity cost cases. 
The performance of the derived guidance law is evaluated in simulations and compared to the optimal-control-based guidance law presented in \cite{drucker2026optimal}. An earlier version of the proposed guidance law was presented by the authors in \cite{drucker2026low}.

The remainder of the paper is organized as follows: In \secref{Model Derivation}, the dynamic model of the problem is presented. In \secref{Problem Formulation and Coordinates Transformation}, the problem is posed, and the coordinate change is performed. In \secref{Optimal Guidance Law}, the game is solved and the terminal speed is constrained. In \secref{Simulations}, the performance of the guidance law is analyzed, followed by the conclusions in \secref{Conclusions}. The normalized dynamics and their use in the conjugate point analysis are presented in the Appendix.

%
\section{Model Derivation} \label{Model Derivation}
\begin{figure}[hbtp] 
\centering
\includegraphics[width=0.55\textwidth]{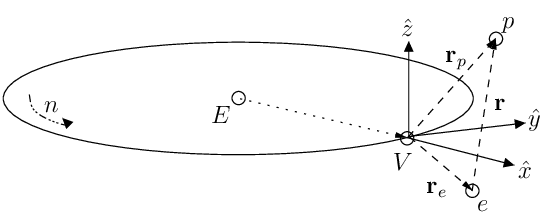}
\caption{The Problem Geometry.}
\label{Problem Geometry}
\end{figure}

We assume two spacecraft near a circular reference orbit around Earth with a mean motion $n$, such that the distance between them is small relative to the reference orbit's semimajor axis. The motion of each spacecraft, relative to a virtual point on the reference orbit (which is close to both spacecraft), can be described by the Clohessy–Wiltshire (CW) equations \cite{clohessy1960terminal}:
\begin{equation} \label{eq: CW equations OC}
\begin{aligned} 
  &\ddot{x}_i - 2n \dot{y}_i - 3n^{2} x_i = u_{x,i}\\
  &\ddot{y}_i + 2n \dot{x}_i = u_{y,i}\\
  &\ddot{z}_i + n^{2} z_i = u_{z,i}
\end{aligned}
\end{equation}
where $x_i,y_i$, and $z_i$ are the three components of the position vector of the spacecraft relative to the virtual point, $i \in \{p,e\}$ denotes the player (pursuer or evader, respectively), $\dot{(\cdot)}$ denotes differentiation with respect to time, and $u_{x,i}, u_{y,i}$, and $u_{z,i}$ are the players' thrust acceleration commands, acting in the $\hat{x},\hat{y}$, and $\hat{z}$ axes, respectively. The moving frame in which the equations are represented is defined as follows: The origin of the frame is at the virtual point, the $\hat{x}$ axis is radially outward, the $\hat{y}$ axis is along the reference orbit, and the $\hat{z}$ axis is along the angular momentum vector of the virtual point, and completes a right-handed coordinates system. 
In \figref{Problem Geometry}, the problem geometry of the CW frame is shown. The Earth’s center, the virtual point, the pursuer, and the evader are denoted by $E, V, p$, and $e$, respectively.

We also assume no external disturbances, such as $J_2$ perturbations, atmospheric drag, or solar radiation pressure.
The EOMs in \eqref{eq: CW equations OC} can be written in a state-space form for each player as follows
\begin{equation} \label{CW equations DG}
\begin{aligned} 
  &\dot \bfx_{p} = \bfA \bfx_{p} + \bfB \bfu_{p} \\
  &\dot \bfx_{e} = \bfA \bfx_{e} + \bfB \bfu_{e}
\end{aligned}
\end{equation}
where the state matrix $\bfA$ and the input matrix $\bfB$ are
\begin{equation} \label{eq: A and B terms OC}
    \bfA = \begin{bmatrix} \mathbf{0}_{3} & \bfI_{3} \\ \bfA_{21} & \bfA_{22} \end{bmatrix}
    ,\quad
    \bfB = \begin{bmatrix} \mathbf{0}_{3} \\ \bfI_{3} \end{bmatrix}
    ,\quad
    \bfA_{21} = \begin{bmatrix} 3n^2 & 0 & 0 \\ 0 & 0 & 0 \\ 0 & 0 & -n^2 \end{bmatrix}
    ,\quad
    \bfA_{22} = \begin{bmatrix} 0 & 2n & 0 \\ -2n & 0 & 0 \\ 0 & 0 & 0 \end{bmatrix}
\end{equation}
and $\mathbf{0}_{m}, \bfI_{m}$ are the zero and the identity matrices with $m \times m$ dimensions, respectively. The corresponding state vector of each player contains the three components of the position and the three components of the velocity of the player relative to the virtual point, respectively
\begin{equation} \label{eq: state vector OC}
    \bfx_i = \begin{bmatrix} {\bfr}_i^{T} & {\bfv}_i^{T} \end{bmatrix}^{T} , \quad  {\bfr}_i=\begin{bmatrix} {x}_i & {y}_i & {z}_i \end{bmatrix}^{T} , \quad  {\bfv}_i=\begin{bmatrix} \dot{{x}}_i & \dot{{y}}_i & \dot{{z}}_i \end{bmatrix}^{T}
\end{equation}
and the control vector of each player contains the three components of the thrust acceleration vector of the player
\begin{equation} \label{eq: control vector OC}
    \bfu_i = \begin{bmatrix}
        {u}_{x,i} & {u}_{y,i} & {u}_{z,i}
    \end{bmatrix}^T
\end{equation}
By subtracting the relative state of the evader from that of the pursuer, we can define a new state vector that describes the motion of the pursuer relative to the evader in the CW frame attached to the virtual point
\begin{equation} \label{new state vector DG}
    \bfx = \bfx_{p} - \bfx_{e} = \begin{bmatrix} {\bfr}^{T} & {\bfv}^{T} \end{bmatrix}^{T}
\end{equation}
and its dynamics are
\begin{equation} \label{eq: mathematical model DG}
    \dot \bfx = \bfA \bfx + \bfB \bfu_{p} - \bfB \bfu_{e}
\end{equation}

\section{Problem Formulation and Coordinates Transformation} \label{Problem Formulation and Coordinates Transformation}
The objective of the pursuer is to achieve a minimum miss distance at the end of the scenario with the required terminal relative speed, while minimizing fuel consumption and accounting for the target's possible evasive maneuver. To do so, a differential-games model is considered, and the problem is formulated with a quadratic cost function, which contains a terminal cost on the states and a running cost on the control effort of each spacecraft
\begin{equation} \label{eq: cost function DG}
    \mathcal{J} = \frac{1}{2}\bfr^{T}(t_{f})\bfQ_{r}\bfr(t_{f}) + \frac{1}{2} \bfv^{T}(t_{f})\bfQ_{v}\bfv(t_{f}) + \frac{1}{2}\int_{0}^{t_{f}} \left(\norm{\bfu_{p}(t)}^{2} - \eta^{2} \norm{\bfu_{e}(t)}^{2} \right) \dd t
\end{equation}
where $\eta>1$ represents the pursuer's maneuver capability relative to that of the evader, $\bfQ_{r}$ and $\bfQ_{v}$ are user-defined diagonal weighting matrices, and $t_f$ is the fixed final time. The optimization problem is defined so that the pursuer minimizes the cost function, while the evader maximizes it.
Note that, unlike the cost function of the optimal-control-based law in \cite{drucker2026optimal}, the desired velocity does not appear explicitly in the cost function of the differential game formulation. Since the two players have opposing objectives, a secondary static optimization, such as minimizing the optimal cost with respect to the constrained velocity, is not applicable, as the cost can be interpreted as either a minimum or a maximum depending on the player's perspective. Instead, the terminal speed is constrained by setting the weighting parameters associated with the terminal velocity in the cost function.
To achieve a sufficiently small miss distance, the weighting matrix $\mathbf{Q}_{r}$ is required to be PD. In contrast, $\mathbf{Q}_{v}$ can be either PD or ND, allowing the terminal speed to be constrained to any desired value. We expect that the resultant final speed will be low (relative to the speed obtained for $\bfQ_v=\mathbf{0}_3$) when $\mathbf{Q}_{v}$ is PD, and high when $\mathbf{Q}_{v}$ is ND.
We define a new state vector using a coordinate transformation, sometimes called "Terminal Projection" \cite{bryson2018applied}
\begin{equation} \label{eq: terminal projection OC}
    \bfz = \begin{bmatrix} \bfz_r^{T} & \bfz_v^{T} \end{bmatrix}^{T} = \bfPhi(t_{f},t)\bfx
\end{equation}
where $\bfPhi(t_{f},t)$ is the state transition matrix associated with the dynamics presented in \eqref{eq: mathematical model DG}. Since the dynamics are autonomous, $\bfPhi(t_{f},t)$ depends on the time-to-go, denoted as $t_{\mathrm{go}}=t_{f}-t$, such that

\begin{equation} \label{eq: Phi term}
    \bfPhi (t_{\mathrm{go}}) = 
    \begin{bmatrix} 
    4-3\,\cos\left(n\,t_{\mathrm{go}}\right)&
    0&
    0&
    \frac{\sin\left(n\,t_{\mathrm{go}}\right)}{n}&
    -\frac{2\,\left[\cos\left(n\,t_{\mathrm{go}}\right)-1\right]}{n}&
    0\\
    6\,\sin\left(n\,t_{\mathrm{go}}\right)-6\,n\,t_{\mathrm{go}}&
    1&
    0&
    \frac{2\,\left[\cos\left(n\,t_{\mathrm{go}}\right)-1\right]}{n}&
    \frac{4\,\sin\left(n\,t_{\mathrm{go}}\right)}{n}-3\,t_{\mathrm{go}}&
    0\\
    0&
    0&
    \cos\left(n\,t_{\mathrm{go}}\right)&
    0&
    0&
    \frac{\sin\left(n\,t_{\mathrm{go}}\right)}{n}\\
    3\,n\,\sin\left(n\,t_{\mathrm{go}}\right)&
    0&
    0&
    \cos\left(n\,t_{\mathrm{go}}\right)&
    2\,\sin\left(n\,t_{\mathrm{go}}\right)&
    0\\
    6\,n\,\left[\cos\left(n\,t_{\mathrm{go}}\right)-1\right]&
    0&
    0&
    -2\,\sin\left(n\,t_{\mathrm{go}}\right)&
    4\,\cos\left(n\,t_{\mathrm{go}}\right)-3&
    0\\
    0&
    0&
    -n\,\sin\left(n\,t_{\mathrm{go}}\right)&
    0&
    0&
    \cos\left(n\,t_{\mathrm{go}}\right)\\
    \end{bmatrix}
\end{equation}
The state transition matrix satisfies the following relations
\begin{subequations} \label{eq: state transition matrix relations OC}
\begin{align}
    &\bfPhi(t_{f},t_{f}) = \bfI \label{eq: identity OC}\\
    &\frac{\dd}{\dd t} \bfPhi(t_{f},t) = -\bfPhi(t_{f},t) \bfA \label{eq: derivative DG}    
\end{align}
\end{subequations}
Substituting $t=t_f$ into \eqref{eq: terminal projection OC}, using \eqref{eq: identity OC}, 
%
%
and substituting 
the result into \eqref{eq: cost function DG} yields the cost function in the new state vector $\bfz$ 
\begin{equation} \label{z cost function DG}
    \mathcal{J} = \frac{1}{2}\bfz^{T}(t_{f})\bfQ\bfz(t_{f}) + \frac{1}{2}\int_{0}^{t_{f}} \left(\norm{\bfu_{p}(t)}^{2} - \eta^{2} \norm{\bfu_{e}(t)}^{2} \right) \dd t
\end{equation}
where
\begin{equation} \label{eq: Q term OC}
 \bfQ = \begin{bmatrix} \bfQ_r & \mathbf{0}_{3} \\ \mathbf{0}_{3} & \bfQ_v \end{bmatrix}, \quad
 \bfQ_r = \begin{bmatrix} q_{x} & 0 & 0 \\ 
    0 & q_{y} & 0 \\
    0 & 0 & q_{z} \end{bmatrix}, \quad
    \bfQ_v = \begin{bmatrix} q_{\dot x} & 0 & 0 \\ 
    0 & q_{\dot y} & 0 \\
    0 & 0 & q_{\dot z}
    \end{bmatrix}
\end{equation}
Taking the time derivative of the new state vector and substituting \eqref{eq: mathematical model DG} and \eqref{eq: derivative DG} yields
\begin{equation} \label{z dot 3 DG}
   \dot \bfz = \Tilde{\bfB} (t_{\mathrm{go}}) \bfu_{p} - \Tilde{\bfB} (t_{\mathrm{go}}) \bfu_{e} \, , \quad \Tilde{\bfB} (t_{\mathrm{go}}) = \bfPhi(t_{\mathrm{go}}) \bfB 
\end{equation}
To improve readability, we will not carry the dependency of $\Tilde{\bfB}(t_{\mathrm{go}})$ on $t_{\mathrm{go}}$ in the rest of this paper. This driftless form of the EOMs, obtained via the coordinate transformation, plays a significant role in solving the differential games problem, as it simplifies the TPBVP.

\section{Optimal Guidance Laws} \label{Optimal Guidance Law} 
\subsection{Optimal Controllers}
The Hamiltonian of the problem is
\begin{equation} \label{hamiltonian DG}
    \mathcal{H} = \frac{1}{2} \left(\bfu_{p}^{T} \bfu_{p} - \eta^{2} \, \bfu_{e}^{T} \bfu_{e} \right) + \bflambda^{T} \left(\Tilde{\bfB} \bfu_{p} - \Tilde{\bfB} \bfu_{e} \right)
\end{equation}
and the corresponding co-state equations are
\begin{subequations} \label{co-states equations DG}
\begin{align}
    &\dot \bflambda^{T} = -\frac{\partial\mathcal{H}}{\partial \bfz} = \mathbf{0}^{T} \label{eq: lambda dot DG}\\
    &\bflambda^{T}(t_{f}) = \frac{\partial}{\partial \bfz(t_{f})} \left[\frac{1}{2}\bfz^{T}(t_{f})\bfQ\bfz(t_{f}) \right]\label{eq: lambda t_f DG} 
\end{align}
\end{subequations}
Since $\bfQ$ is symmetric, the solution of \eqref{eq: lambda dot DG} with the boundary condition in \eqref{eq: lambda t_f DG} is
\begin{equation} \label{lambda DG}
    \bflambda(t) \equiv \bflambda(t_{f}) = \bfQ\bfz(t_{f})
\end{equation}
The pursuer's optimal controller $\bfu_{p}^{*}$ and the evader's optimal controller $\bfu_{e}^{*}$ are given by
\begin{equation} \label{arg min H DG}
    \bfu_{p}^{*} = \arg \min_{\bfu_{p}} \mathcal{H} \, , \quad \bfu_{e}^{*} = \arg \max_{\bfu_{e}} \mathcal{H}
\end{equation}
Since the controllers are assumed unbounded, the following first-order necessary conditions have to be satisfied
\begin{equation} \label{dH/du DG}
    \frac{\partial \mathcal{H}}{\partial \bfu_{p}} = \bfu_{p}^{T} +  \bflambda^{T}(t)\Tilde{\bfB} = \mathbf{0}^{T}
    \, , \quad
    \frac{\partial \mathcal{H}}{\partial \bfu_{e}} = -\eta^2 \bfu_{e}^{T} - \bflambda^{T}(t)\Tilde{\bfB} = \mathbf{0}^{T}
\end{equation}
and by substituting \eqref{lambda DG} into \eqref{dH/du DG} we get the optimal open-loop controllers 
\begin{equation} \label{u star open DG}
    \bfu_{p}^{*} = -\Tilde{\bfB}^{T} \bfQ\bfz(t_{f})
    \, , \quad
    \bfu_{e}^{*} = - \frac{1}{\eta^{2}} \Tilde{\bfB}^{T} \bfQ\bfz(t_{f})
\end{equation}
To get the optimal controllers, an explicit expression for $\bfz(t_{f})$ is needed. Substituting \eqref{u star open DG} into \eqref{z dot 3 DG} and integrating from $t$ to $t_{f}$ yields
\begin{equation} \label{integrating z dot DG}
    \bfz(t_{f}) - \bfz(t) = - \left(1 - \frac{1}{\eta^{2}}\right) \int_{t}^{t_{f}} \Tilde{\bfB}(\mathbf{\tau}) \Tilde{\bfB}^{T}(\mathbf{\tau}) \,\dd \mathbf{\tau} \bfQ\bfz(t_{f}) \quad \Longrightarrow \quad \bfz(t) = \bfP \bfz(t_{f})
\end{equation}
where
\begin{equation} \label{P tilde M chal_B and chal_C DG}
    \bfP = \bfI_6 + \gamma \bfchalB \bfQ \, , \quad \gamma = 1 - \frac{1}{\eta^{2}}>0 \, , \quad \bfchalB = \int_{t}^{t_{f}} \Tilde{\bfB}(\mathbf{\tau}) \Tilde{\bfB}^{T}(\mathbf{\tau}) \,\dd \mathbf{\tau}
\end{equation}
So if the matrix $\bfP$ is invertible, then $\bfz(t_{f}) = \bfP^{-1} \bfz(t)$ and the closed-loop optimal controllers of the pursuer and the evader are, respectively
\begin{equation} \label{eq: u v star closed DG}
    \bfu_{p}^{*} = -\Tilde{\bfB}^{T} \bfQ \bfP^{-1} \bfz(t) \, , \quad \bfu_{e}^{*} = -\frac{1}{\eta^{2}} \Tilde{\bfB}^{T} \bfQ \bfP^{-1} \bfz(t)
\end{equation}

Let us examine the case where $\eta \rightarrow \infty$. In this case, the maneuver capability of the pursuer is much greater than that of the evader, such that the evader does not maneuver. From \eqref{eq: u v star closed DG} it is evident that indeed in this case $\bfu_{e}^{*} \equiv \mathbf{0}$, and from \eqref{P tilde M chal_B and chal_C DG} that $\gamma \rightarrow 1$ and therefore $\bfu_{p}^{*}$ converges to the classic optimal-control-based solution obtained in \cite{drucker2026optimal}.

\subsection{Conjugate Point Analysis}
In order to get the optimal controllers, the matrix $\bfP$ must be invertible for any $0\le t <t_f$ \cite{ben1998advances}. The matrix $\bfP$ can also be written as
\begin{equation} \label{P tilde rewritten DG}
    \bfP = \left(\bfQ^{-1} + \gamma \bfchalB \right)\bfQ = \left(\hat{\bfQ}^{-1} + \bfchalB \right) \hat{\bfQ} \, , \quad \hat{\bfQ}=\gamma \bfQ
\end{equation}
Let us examine two cases for $\bfQ_v$; in the first, $\bfQ_v$ is PD, and in the second, it is ND. Both cases are necessary to constrain the terminal speed to any desired value.

\textit{Case I:} When $\bfQ_v$ is PD, $\bfQ$ is also PD, and so is $\hat{\bfQ}$ and its inverse. Since $\hat{\bfQ}^{-1}$ is PD, and $\bfchalB$ is a definite integral of a Positive-Semi-Definite (PSD) matrix and therefore it is also a PSD matrix, the matrix $\left(\hat{\bfQ}^{-1} + \bfchalB \right)$ is a PD matrix and therefore it is invertible for any time. So, since $\hat{\bfQ}$ is invertible, and the product of two invertible matrices is also invertible, the matrix $\bfP$ is invertible.

\textit{Case II:} The case where $\bfQ_v$ is ND requires a more detailed analysis, similar to the one in \cite{gill1969optimal,ben1996linear}. This analysis is necessary because the cost function’s components in this case are non-positive-definite, and a conjugate point may appear.
Since $\hat{\bfQ}$ is invertible, the term $\left(\hat{\bfQ}^{-1} + \bfchalB \right)$ has to be also invertible for the matrix $\bfP$ to be invertible. The matrix $\bfchalB$ can also be written in a block matrix form as in \eqref{eq: Q term OC}, such that
\begin{equation}
    {\hat{\bfQ}}^{-1} + \bfchalB = \begin{bmatrix} {\hat{\bfQ}}_r^{-1} + \bfchalB_{11} & \bfchalB_{12} \\ \bfchalB_{21} & {\hat{\bfQ}}_v^{-1} + \bfchalB_{22} \end{bmatrix}
\end{equation}
where $\hat{\bfQ}_r=\gamma \bfQ_r$ and $\hat{\bfQ}_v=\gamma \bfQ_v$. Since $\bfchalB_{11}$ is a definite integral of a PSD matrix and therefore it is also a PSD matrix and $\hat{\bfQ}_r$ is a PD matrix, the matrix $\left( {\hat{\bfQ}}_r^{-1} + \bfchalB_{11} \right)$ is invertible and the determinant of $\left({\hat{\bfQ}}^{-1} + \bfchalB \right)$ can be obtained by
\begin{equation} \label{eq: determinant 1}
    \det \left( {\hat{\bfQ}}^{-1} + \bfchalB \right) = \det \left( {\hat{\bfQ}}_r^{-1} + \bfchalB_{11} \right) \det \left[ {\hat{\bfQ}}_v^{-1} + \bfchalB_{22} - \bfchalB_{21} \left( {\hat{\bfQ}}_r^{-1} + \bfchalB_{11} \right)^{-1} \bfchalB_{12} \right]
\end{equation}
Let us now examine the case where $q_{i} \rightarrow \infty \, ; \, i \in \{x, y, z\}$. This case is equivalent to the case where the hard constraint $\bfr(t_{f})=\mathbf{0}$ is required. In this case ${\hat{\bfQ}}_r^{-1}$ vanishes and \eqref{eq: determinant 1} simplifies to
\begin{equation} \label{eq: determinant 2}
    \det \left( {\hat{\bfQ}}^{-1} + \bfchalB \right) = \det \left(\bfchalB_{11} \right) \det \left( {\hat{\bfQ}}_v^{-1} + \bfS \right) \, , \quad \bfS = \bfchalB_{22} - \bfchalB_{21} \bfchalB_{11}^{-1} \bfchalB_{12}
\end{equation}
Let us also consider the following case 
\begin{equation} \label{eq: q_i 1/b}
    q_{i} = -\frac{1}{\gamma b} \, ; \quad i \in \{\dot x, \dot y, \dot z\} , \, b \neq 0
\end{equation}
where $b$ is a user-defined parameter. In this case ${\hat{\bfQ}}_v = -\frac{1}{b} \, \bfI_3$, so for $b<0$ the matrix ${\bfQ}_v$ is PD and for $b>0$ it is ND. Therefore, \eqref{eq: determinant 2} simplifies to
\begin{equation} \label{eq: determinant 3}
    \det \left( {\hat{\bfQ}}^{-1} + \bfchalB \right) = \det \left(\bfchalB_{11} \right) \det \left( \bfS - b \bfI \right)
\end{equation}
The matrix $\bfchalB_{11}$ is PD for all $t_{\mathrm{go}} > 0$, so its determinant is positive. The other determinant will be zero when $b$ equals one of the eigenvalues of $\bfS$ at a certain time throughout the mission.

To improve readability, the analysis of the behavior of the eigenvalues of $\bfS$ as a function of $t_{\mathrm{go}}$, in normalized coordinates to cover all altitudes, is presented in the Appendix. From the Appendix, it is evident that the eigenvalues of $\bfS$ are zero at $t_{\mathrm{go}}=0$ and are \emph{monotonically increasing} with $t_{\mathrm{go}}$. Therefore, if $b$ is either higher than all the eigenvalues of $\bfS$ at $t_{\mathrm{go}}=t_f$ or negative, then the matrix $\bfP$ is invertible for any $0\leq t<t_f$. As a result, the solution does not contain a conjugate point, and its optimality is guaranteed. This condition does not depend on the initial conditions, reference orbit, %
and maneuverability ratio ($\eta$). Moreover, this condition has to be tested only at $t_{\mathrm{go}}=t_f$, yet it holds for any $t_{\mathrm{go}}$.  

\subsection{Constraining the Terminal Speed} \label{Constrained Terminal Speed}
Assuming that $q_{i} \rightarrow \infty \, ; \, i \in \{x, y, z\}$, and that $b$ is chosen such that the matrix ${\bfP}$ is not singular throughout the scenario, the final relative velocity can be obtained by
\begin{equation}
    \bfv(t_f) = \bfB^T \bfx(t_f) = \bfB^T \bfz(t_f) = \bfB^T {\bfP}^{-1} \bfz
\end{equation}
where
\begin{equation} \label{eq: P inverse OC}
    \bfB^T {\bfP}^{-1} = \begin{bmatrix} b\left( \bfS - b \bfI \right)^{-1} \bfchalB_{21} \bfchalB_{11}^{-1} \, & \, -b\left( \bfS - b \bfI \right)^{-1} \end{bmatrix}
\end{equation}
Therefore, the final relative velocity is
\begin{equation}
    \bfv(t_f) = b \left( \bfS - b \bfI \right)^{-1} \bfq , \quad \bfq = \bfchalB_{21} \bfchalB_{11}^{-1} \bfz_r - \bfz_v
\end{equation}
So the magnitude of the final relative velocity can be obtained by
\begin{equation} \label{eq: v_c^2}
    \norm{\bfv(t_f)}^2 = \bfv^T(t_f) \bfv(t_f) = b^2 \bfq^T \left( \bfS - b \bfI \right)^{-2} \bfq = v_c^2
\end{equation}
where $v_c$ denotes the desired terminal speed, which is called the command speed. Diagonalizing the matrix $\left( \bfS - b \bfI \right)$
\begin{equation} \label{eq: diagonalizing}
    \bfS - b \bfI = \bfV \left( \bfLambda - b \bfI \right) \bfV^{-1}
\end{equation}
where $\bfLambda$ is a diagonal $3 \times 3$ matrix of the eigenvalues of the matrix $\bfS$, and $\bfV$ is a $3 \times 3$ matrix whose columns are the corresponding unit eigenvectors of $\bfS$. Since $\bfS$ is a symmetric matrix, the matrix $\bfV$ is an orthonormal matrix, i.e., $\bfV^{-1}=\bfV^T$. Substituting \eqref{eq: diagonalizing} into \eqref{eq: v_c^2} yields
\begin{equation} \label{eq: v_c^2 2}
    b^2 \bfa^T \left( \bfLambda - b \bfI \right)^{-2} \bfa = v_c^2
\end{equation}
where $\bfa = \bfV^T \bfq$. \eqref{eq: v_c^2 2} can be written as
\begin{equation} \label{eq: v_c^2 3}
    b^2 \sum_{i=1}^{3} \frac{a_i^2}{(\lambda_i-b)^2} = v_c^2
\end{equation}
where $\lambda_i$ and $a_i \, ; \, i \in \{1,2,3\}$ are the eigenvalues of $\bfS$ and the components of the vector $\bfa$, respectively. \eqref{eq: v_c^2 3} can be simplified to
\begin{equation} \label{eq: mu star 4}
    v_c^2 (\lambda_1-b)^2 (\lambda_2-b)^2 (\lambda_3-b)^2 - b^2 \left[ a_1^2 (\lambda_2-b)^2 (\lambda_3-b)^2 + a_2^2 (\lambda_1-b)^2 (\lambda_3-b)^2 + a_3^2 (\lambda_1-b)^2 (\lambda_2-b)^2 \right] = 0
\end{equation}
This is a sixth-order algebraic equation in the user-defined parameter $ b$ that can be solved for $b$, given the state vector $\bfx$ and the time-to-go.

To prove that any desired speed can be obtained without encountering a conjugate point, let us denote $ f(b) = \norm{\bfv(t_f)}$. From \eqref{eq: v_c^2} and \eqref{eq: v_c^2 3} it is evident that $f(b)$ is continuous for $b \in (\lambda_{\mathrm{max}},\infty)$, where $\lambda_{\mathrm{max}}$ is the highest eigenvalue of $\bfS$, and that
\begin{equation} \label{eq: unconstrained speed}
    \lim_{\quad b\to\lambda_{\mathrm{max}}^+} f(b) = \infty \, , \quad \lim_{b\to \infty} f(b) = \norm{\bfa}
\end{equation}
So, according to the intermediate value theorem, there exists some $b^*>\lambda_{\mathrm{max}}$ such that $f(b^*) = \norm{\bfv(t_f)} = v_c$, where $\norm{\bfa}<v_c$. Such a solution satisfies the constraint and does not yield a conjugate point, as proven in the conjugate point analysis.

In addition, because the eigenvalues of $\bfS$ are non-negative $\forall t_{\mathrm{go}} \geq 0$ (as shown in the Appendix), $f(b)$ is also continuous for $b \in (-\infty,0)$, and from \eqref{eq: v_c^2} and \eqref{eq: v_c^2 3} it is evident that
\begin{equation} \label{eq: unconstrained speed 2}
    \lim_{\, b\to -\infty} f(b) = \norm{\bfa} \, , \quad \lim_{b \to 0} f(b) = 0
\end{equation}
So, according to the intermediate value theorem, there exists some $b^*<0$ such that $f(b^*) = \norm{\bfv(t_f)} = v_c$, where $0<v_c<\norm{\bfa}$. Such a solution satisfies the constraint and does not yield a conjugate point, as proven in the conjugate point analysis.
Therefore, a solution that satisfies the terminal-speed constraint without a conjugate point can be obtained for any $v_c \ge 0$.

We now examine two special cases of $b$. In the first case, $b \rightarrow 0^-$, so from \eqref{eq: q_i 1/b} it is evident that $q_i \rightarrow \infty \, ; \, i \in \{\dot x, \dot y, \dot z\}$. Therefore, this case is equivalent to the case where the hard constraint $\bfv(t_{f})=\mathbf{0}$ is also required, i.e., a perfect rendezvous.
The second case is when $b \rightarrow \infty$. From \eqref{eq: q_i 1/b} it is evident that $q_i \rightarrow 0 \, ; \, i \in \{\dot x, \dot y, \dot z\}$ in this case, so $\bfQ_v=\mathbf{0}_3$ and the terminal velocity does not appear in the cost function. Therefore, this case is equivalent to the case where the terminal velocity is unconstrained. Note that the same result can be obtained for $b \rightarrow -\infty$. The unconstrained terminal speed can be obtained using \eqref{eq: v_c^2 3} and its value is $\norm{\bfv(t_f)}=\norm{\bfa}=\norm{\bfq}$.

Let us consider the case where the desired final relative velocity ($v_c$) is relatively high, such that \eqref{eq: v_c^2 3} becomes three separated equations, in which one component of the equation is much higher than the other two in each equation
\begin{equation} \label{eq: b star approx}
    b^2 \sum_{i=1}^{3} \frac{a_i^2}{(\lambda_i-b)^2} \approx b^2 \frac{a_k^2}{(\lambda_k-b)^2} = v_c^2 \, , \quad k \in \{1, 2, 3\}
\end{equation}
Each equation is a quadratic equation of the form
\begin{equation} \label{eq: b star approx 2}
    \left( v_c^2 - a_k^2 \right) b^2 - 2 \lambda_k v_c^2 b + \lambda_k^2 v_c^2 = 0
\end{equation}
and the solutions satisfy
%
\begin{equation} \label{eq: b star approx 3}
    b = \frac{\lambda_k v_c^2 \pm a_k \lambda_k v_c}{v_c^2 - a_k^2} = \frac{v_c}{v_c \pm a_k} \lambda_k 
\end{equation}
%
%
Since the command speed is assumed to be relatively high, $b$ has to be higher than all the eigenvalues of $\bfS$ to avoid a conjugate point. The approximate solution for $b$, which satisfies this condition, is the one that is slightly higher than the maximal eigenvalue
\begin{equation} \label{eq: b star approx 4}
    b^* = \frac{v_c}{v_c - |a_{\mathrm{max}}|} \lambda_{\mathrm{max}}
\end{equation}
where $a_{\mathrm{max}}$ is the component of $\bfa$ associated with $\lambda_{\mathrm{max}}$. This solution for $b$ can be used as an initial guess for a numeric solution of \eqref{eq: mu star 4}, or even as the solution itself if the relatively high command speed assumption is valid.

\section{Simulations} \label{Simulations}
\subsection{Simulations Scenario}
A mission scenario with the following parameters was considered: The reference-orbit's height was $h = 750$ (km), the scenario duration was $t_{f} = 2T$ where $T$ is the period of the reference orbit, the initial position of the pursuer relative to the evader was $\bfr_{0} = \begin{bmatrix} 0 & -5 & -1\end{bmatrix}^{T} (\text{km})$, and the initial velocity of the pursuer relative to the evader was $\bfv_{0} = \begin{bmatrix} -1 & 2 & 1\end{bmatrix}^{T} \left(\frac{\text{m}}{\text{s}}\right)$. The constrained terminal speed was $v_c=15$ (m/s), and the pursuer's maneuver capability relative to the evader's maneuver capability was $\eta=2$.
The parameter $b$ was numerically calculated from \eqref{eq: mu star 4}, in a closed-loop, using the MATLAB \textit{roots} function, which computes the eigenvalues of the corresponding companion matrix \cite{edelman1995polynomial}. However, $b$ was kept constant for the last 30 seconds of the scenario to avoid controller chatter at the end of the simulations. The weighting matrix $\bfQ_r$ was chosen to be $\bfQ_r = \frac{n^2}{\gamma|b|} \, \bfI_3 \cdot 10^8$. The ODEs presented in \eqref{eq: CW equations OC} were solved using the MATLAB \textit{ode45} function, which is based on an explicit fourth-order Runge-Kutta formula \cite{shampine1997matlab}. In all simulations, the thrust acceleration of the pursuer was limited by $2 \cdot 10^{-6} \left( \frac{\text{km}}{\text{s}^{2}} \right)$ and that of the evader by $1 \cdot 10^{-6} \left( \frac{\text{km}}{\text{s}^{2}} \right)$, to represent realistic limitations of the propulsion system.

\subsection{Simulations Results}
Two simulations were performed for the guidance law presented in \eqref{eq: u v star closed DG}, each with a different target strategy; in the first the target was optimally evading (OET), using its optimal strategy (also presented in \eqref{eq: u v star closed DG}), and in the second the target did not apply any control throughout the mission (NMT). In \tref{Simulation Results, High Terminal Speed}, the miss distance, terminal speed miss, control effort of the pursuer, and the total running cost obtained in the simulations for both targets' strategies are presented. The guidance law performs better in terms of miss distance and terminal speed miss in the OET case. The NMT case has higher misses because the target does not maneuver as the guidance law expects. Yet, the miss distance and the terminal speed miss in the NMT case are small. In the OET case, the pursuer's and the total running costs are higher because the target uses its optimal strategy. When the target deviates from it, the cost decreases, as in the NMT case (in this case, the pursuer's running cost and the total running cost coincide because the target's control effort in the cost function is nullified).
\begin{table}[hbtp]
\centering
\caption{Simulation Results for Two Target Strategies. \label{Simulation Results, High Terminal Speed}}
\begin{tabular}{@{}ccc@{}}
\toprule \toprule
Target Strategy & Optimally Evading & Non-Maneuvering \\
\midrule 
Miss Distance $\left( \text{m} \right)$                                      & $7.194 \times 10^{-4}$ & $2.557 \times 10^{-2}$ \\
Speed Miss $\left(\frac{\text{m}}{\text{s}} \right)$                         & $1.468 \times 10^{-4}$ & $1.558\times 10^{-2}$ \\ 
Pursuer's Running Cost $\left( \frac{\text{km}^{2}}{\text{s}^{3}} \right)$ & $7.6631 \times 10^{-9}$ & $4.5918 \times 10^{-9}$ \\
Total Running Cost $\left( \frac{\text{km}^{2}}{\text{s}^{3}} \right)$     & $5.7473 \times 10^{-9}$ & $4.5918 \times 10^{-9}$ \\ 
\bottomrule \bottomrule
\end{tabular}
\end{table}

\begin{figure}[hbtp]
\centering
\includegraphics[width=0.75\textwidth]{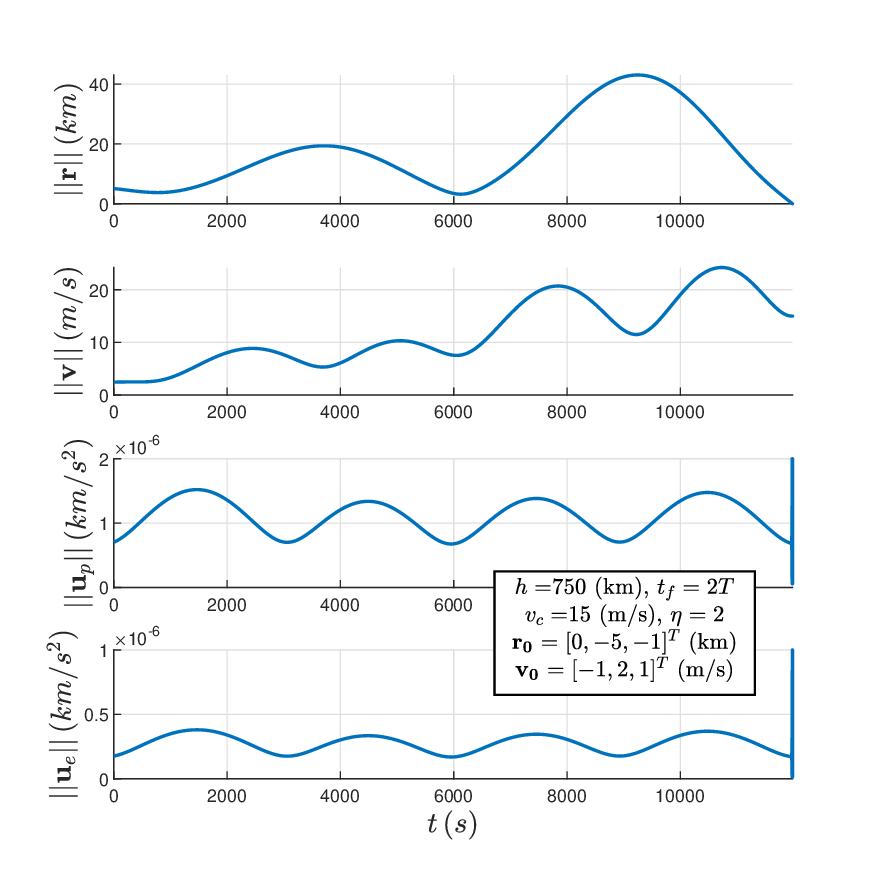}
\caption{\label{fig: NPDC_OMT_HS_norms}  Time Histories of the Magnitudes of the Relative Distance, Relative Speed, and Thrust Accelerations - Optimally Evading Target.}
\end{figure}

\begin{figure}[hbtp]
\centering
\includegraphics[width=0.75\textwidth]{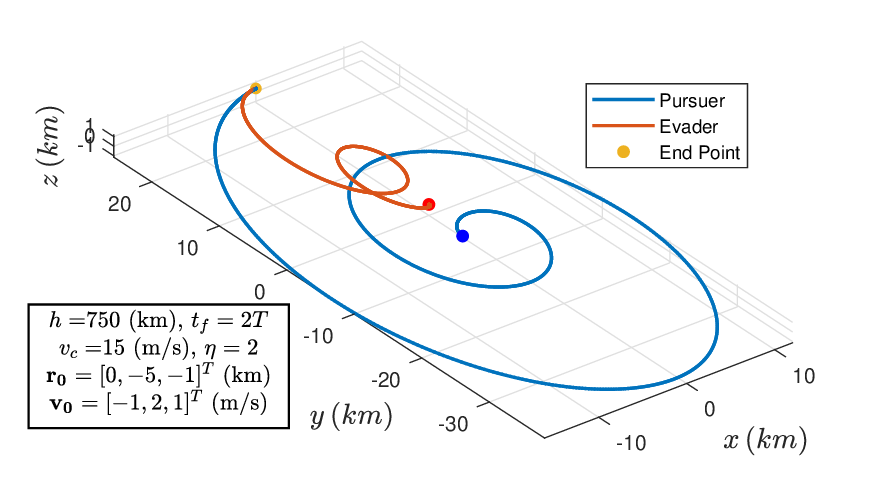}
\caption{\label{fig: NPDC_OMT_HS_trajectory}  Relative Trajectories - Optimally Evading Target.}
\end{figure}

\begin{figure}[hbtp]
\centering
\includegraphics[width=0.75\textwidth]{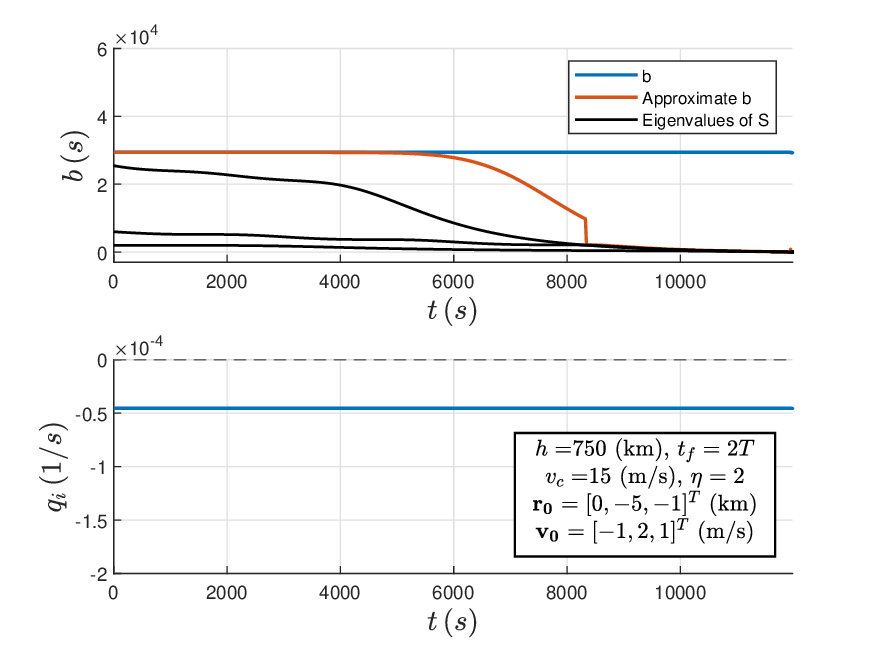}
\caption{\label{fig: NPDC_OMT_HS_bq}  Time Histories of the Terminal Speed's Weighting Parameters - Optimally Evading Target.}
\end{figure}

\begin{figure}[hbtp]
\centering
\includegraphics[width=0.75\textwidth]{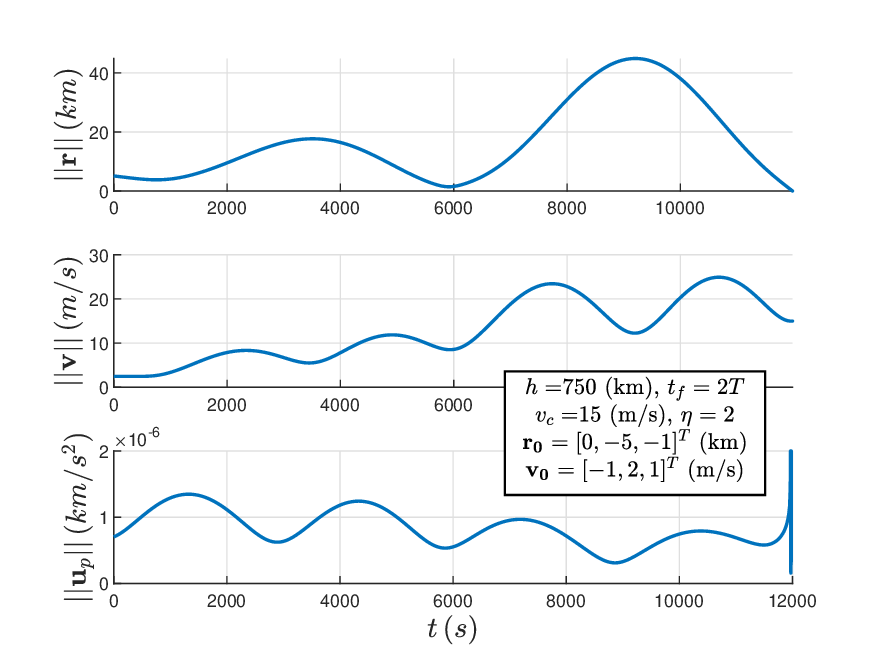}
\caption{\label{fig: NPDC_NMT_HS_norms}  Time Histories of the Magnitudes of the Relative Distance, Relative Speed, and Thrust Acceleration - Non-Maneuvering Target.}
\end{figure}

\begin{figure}[hbtp]
\centering
\includegraphics[width=0.75\textwidth]{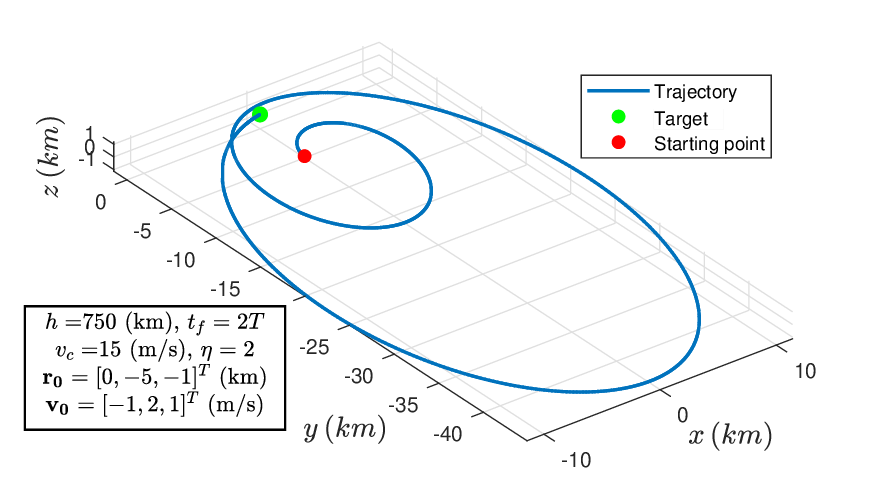}
\caption{\label{fig: NPDC_NMT_HS_trajectory}  Relative Trajectories - Non-Maneuvering Target.}
\end{figure}

\begin{figure}[hbtp]
\centering
\includegraphics[width=0.75\textwidth]{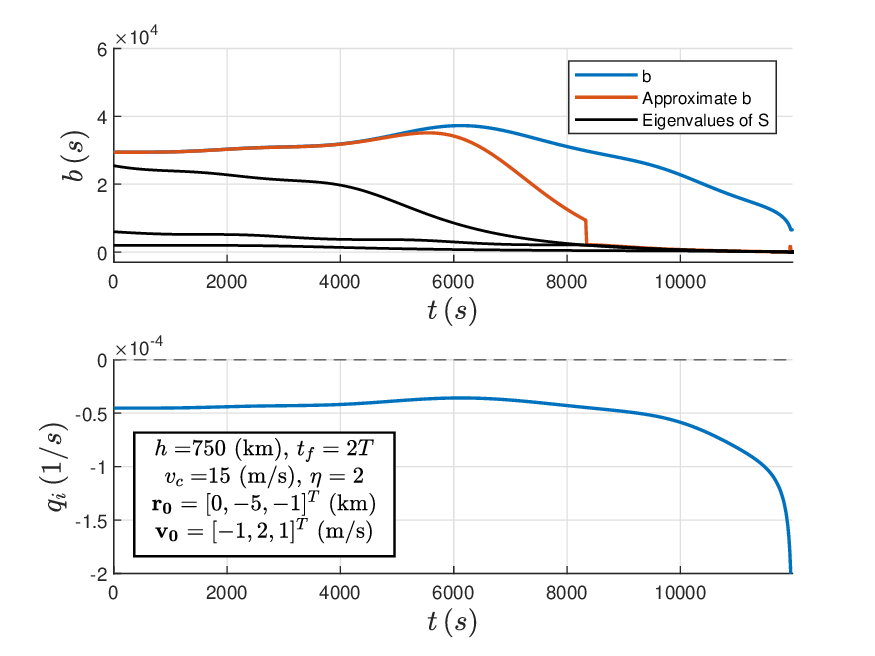}
\caption{\label{fig: NPDC_NMT_HS_bq}  Time Histories of the Terminal Speed's Weighting Parameters - Non-Maneuvering Target.}
\end{figure}

In \figref{fig: NPDC_OMT_HS_norms}, the time histories of the relative distance, relative speed, and the thrust acceleration norms of both players are presented for the OET case. In \figref{fig: NPDC_OMT_HS_trajectory}, the trajectories of both players, relative to the origin, are presented. It is assumed that the evading spacecraft starts from the origin, with zero relative velocity. In \figref{fig: NPDC_OMT_HS_bq}, the time histories of the parameter $b$, together with the eigenvalues of $\bfS$ and the approximation of $b$, calculated using \eqref{eq: b star approx 4}, and the weighting parameter $q_{i} \, ; \, i \in \{\dot x, \dot y, \dot z\}$, are presented for the OET case. Since the assumption of an OET is not violated, both parameters remain constant throughout the mission. The analytical approximation of $b$ initially matches the exact numerical solution because the terminal speed of the unconstrained case (which can be obtained from \eqref{eq: unconstrained speed}) for this simulation scenario was $v=2.260$ (m/s) at the beginning of the scenario. However, this approximation eventually deviates from the exact solution. This happens because the unconstrained speed increases with time, so the relatively high constrained terminal speed assumption is violated. In addition, since the constrained terminal speed is relatively high, the solution obtained for $b$, which avoids a conjugate point, is positive, and the speed weighting parameter $q_{i}$ is therefore negative, as expected.
%
In \figref{fig: NPDC_NMT_HS_norms}, the time histories of the relative distance, relative speed, and the thrust acceleration norm of the pursuer are presented for the NMT case. Since the target does not maneuver as assumed in the guidance law, a relatively high thrust acceleration is required at the end of the scenario. In \figref{fig: NPDC_NMT_HS_trajectory}, the trajectory of the pursuer relative to the target (located at the origin) is presented. In \figref{fig: NPDC_NMT_HS_bq}, the time histories of the parameter $b$, together with the eigenvalues of $\bfS$, the approximation of $b$, and the weighting parameter $q_{i} \, ; \, i \in \{\dot x, \dot y, \dot z\}$, are presented for the NMT case. Since the OET assumption is violated in this case, neither parameter remains constant. Yet, the solutions obtained for $b$ remain higher than the eigenvalues of $\bfS$ throughout the mission, and therefore a conjugate point is avoided. The analytical approximation of $b$ initially matches the exact numerical solution, but eventually deviates for the same reason as in the OET case.

%

%

\subsection{Comparison to the Optimal Control Guidance Law}
The differential-games-based guidance law (denoted as DGGL) presented in \eqref{eq: u v star closed DG} and the optimal-control-based guidance law (denoted as OGL) presented in \cite{drucker2026optimal} (which can also be obtained from \eqref{eq: u v star closed DG} by setting $\eta \rightarrow \infty$) were tested in simulations with the same initial conditions, but with several values of $\eta$, in two cases: an NMT case and an OET case. The chosen command speed for both guidance laws was $v_c=15$ (m/s).

\begin{figure}[hbtp]
\centering
\includegraphics[width=0.75\textwidth]{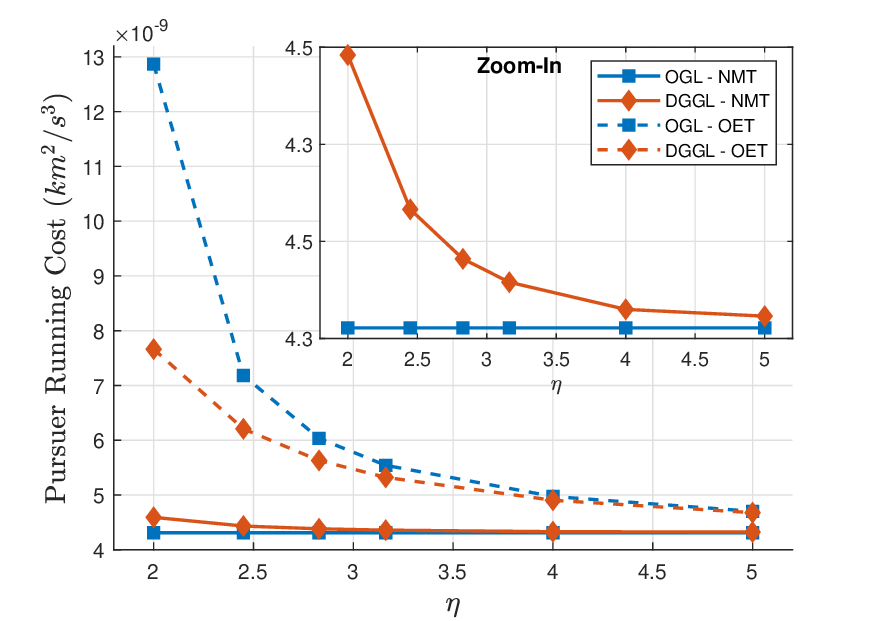}
\caption{\label{fig: costs_graph_2}  Pursuer's Running Costs Comparison, NMT and OET cases.}
\end{figure}

In \figref{fig: costs_graph_2}, the running costs obtained by each guidance law for the pursuer in the simulations are presented for both cases. In the NMT case, the costs obtained by the OGL are the same in all simulations in this case, because it does not depend on $\eta$ and the target does not maneuver. The costs obtained by the DGGL are higher than those of the OGL, since the OGL was specifically derived for an NMT case. However, it is evident that as $\eta$ increases, the cost of the DGGL decreases and approaches the cost obtained by the OGL.

\begin{figure}[hbtp]
\centering
\includegraphics[width=0.75\textwidth]{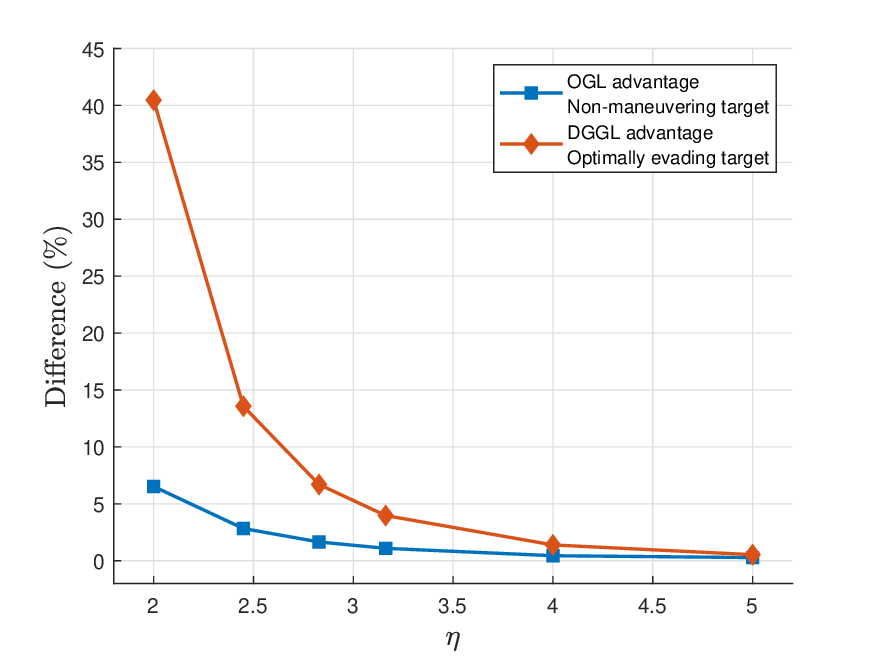}
\caption{\label{fig: costs_graph_1}  Differences in the Pursuer's Running Costs, NMT and OET cases.}
\end{figure}

In the OET case, the costs obtained by the OGL are higher than those of the DGGL, because the DGGL was specifically derived for an OET case. As in the NMT case, the difference in the cost decreases as $\eta$ increases. Yet, for each tested value of $\eta$, the difference in cost of the DGGL in the OET case, i.e., its advantage over the OGL, is much greater than that of the OGL in the NMT case. This difference is also presented in \figref{fig: costs_graph_1}, and it emphasizes the robustness advantage of the DGGL over the OGL.



\section{Conclusions} \label{Conclusions}
This paper investigated a PE differential game for two low-thrust spacecraft. Based on the CW equations, the problem was formulated as a linear-quadratic two-sided optimization problem, where one player aims to minimize the cost and the other to maximize it. The relative position and velocity at the end of the maneuver were softly constrained in the cost function, along with a running cost on each player's control effort. Optimal guidance laws were derived for both players, and a conjugate point analysis was performed to verify the optimality of the solution. The weighting parameters of the terminal velocity in the cost function were selected to constrain the terminal speed to a user-defined value. The performance of the proposed guidance laws was evaluated in simulations with both optimally evading and non-maneuvering targets. It was demonstrated that the proposed pursuer's guidance law can achieve successful engagement while satisfying the constraint. It was also shown that the proposed guidance law outperforms the optimal-control-based state-of-the-art guidance law, in terms of fuel consumption, when facing an optimally evading target. Therefore, the proposed guidance law is better suited to scenarios involving unpredictable targets.

\section*{Appendix: Conjugate Point Analysis in Normalized Dynamics} \label{Appendix}
To prove that conjugate points can be avoided for any $n$, the relative position components in \eqref{eq: CW equations OC} can be normalized by the semimajor axis of the reference orbit $a$, and the angular velocities by $n$, such that the normalized CW equations become \cite{alfriend2009spacecraft}
\begin{equation} \label{eq: normalized CW equations OC}
\begin{aligned} 
  &{\Bar{x}}''_i - 2 {\Bar{y}}'_i - 3 \Bar{x}_i = \Bar{u}_{x,i}\\
  &{\Bar{y}}''_i + 2 {\Bar{x}}'_i = \Bar{u}_{y,i}\\
  &{\Bar{z}}''_i + \Bar{z}_i = \Bar{u}_{z,i}
\end{aligned}
\end{equation}
where $\Bar{(\cdot)}$ denotes a normalized variable, $i \in \{p,e\}$ denotes the player (pursuer or evader, respectively), and ${(\cdot)}'$ denotes differentiation with respect to the normalized time 
\begin{equation}
    \begin{aligned}
    &\Bar{t}=n \, t \quad \Longrightarrow \quad \frac{\dd}{\dd \Bar{t}} (\cdot) = \frac{1}{n} \frac{\dd}{\dd t} (\cdot) \\
    &\Bar{x}_i = \frac{x_i}{a} \, , \quad \Bar{y}_i = \frac{y_i}{a} \, , \quad \Bar{y}_i = \frac{x_i}{a} \\
    &{\Bar{x}}'_i = \frac{\dot{x}_i}{na} \, , \quad {\Bar{y}}'_i = \frac{\dot{y}_i}{na} \, , \quad {\Bar{z}}'_i = \frac{\dot{z}_i}{na} \\
    &\Bar{u}_{x,i} = \frac{u_{x,i}}{n^2a} \, , \quad \Bar{u}_{x,i} = \frac{u_{x,i}}{n^2a} \, , \quad \Bar{u}_{x,i} = \frac{u_{x,i}}{n^2a} \\
    \end{aligned}
\end{equation}
The dynamics of the game can be represented in a state-space block-matrix form, with the normalized relative coordinates, as follows
\begin{equation} \label{eq: normalized mathematical model DG}
    \Bar{\bfx}' = \Bar{\bfA} \Bar{\bfx} + \bfB \Bar{\bfu}_{p} - \bfB \Bar{\bfu}_{e}
\end{equation}
where the normalized state matrix $\bar{\bfA}$ is
\begin{equation} \label{eq: normalized A and B terms OC}
    \Bar{\bfA} = \begin{bmatrix} \mathbf{0}_{3} & \bfI_{3} \\ \Bar{\bfA}_{21} & \Bar{\bfA}_{22} \end{bmatrix}
    ,\quad
    \Bar{\bfA}_{21} = \begin{bmatrix} 3 & 0 & 0 \\ 0 & 0 & 0 \\ 0 & 0 & -1 \end{bmatrix}
    ,\quad
    \Bar{\bfA}_{22} = \begin{bmatrix} 0 & 2 & 0 \\ -2 & 0 & 0 \\ 0 & 0 & 0 \end{bmatrix}
\end{equation}
and the corresponding normalized state and control vectors for each player are defined as in \eqref{eq: state vector OC} and \eqref{eq: control vector OC}, respectively. The corresponding normalized state transition matrix is 
\begin{equation} \label{eq: normalized Phi term}
    \Bar{\bfPhi} (\Bar{t}_{\mathrm{go}}) = 
    \begin{bmatrix} 
    4-3\cos\left(\Bar{t}_{\mathrm{go}}\right)&
    0&
    0&
    \sin\left(\Bar{t}_{\mathrm{go}}\right)&
    -2\left[\cos\left(\Bar{t}_{\mathrm{go}}\right)-1\right]&
    0\\
    6\left[\sin\left(\Bar{t}_{\mathrm{go}}\right)-\Bar{t}_{\mathrm{go}}\right]&
    1&
    0&
    2\left[\cos\left(\Bar{t}_{\mathrm{go}}\right)-1\right]&
    4\sin\left(\Bar{t}_{\mathrm{go}}\right)-3\Bar{t}_{\mathrm{go}}&
    0\\
    0&
    0&
    \cos\left(\Bar{t}_{\mathrm{go}}\right)&
    0&
    0&
    \sin\left(\Bar{t}_{\mathrm{go}}\right)\\
    3\sin\left(\Bar{t}_{\mathrm{go}}\right)&
    0&
    0&
    \cos\left(\Bar{t}_{\mathrm{go}}\right)&
    2\sin\left(\Bar{t}_{\mathrm{go}}\right)&
    0\\
    6\left[\cos\left(\Bar{t}_{\mathrm{go}}\right)-1\right]&
    0&
    0&
    -2\sin\left(\Bar{t}_{\mathrm{go}}\right)&
    4\cos\left(\Bar{t}_{\mathrm{go}}\right)-3&
    0\\
    0&
    0&
    -\sin\left(\Bar{t}_{\mathrm{go}}\right)&
    0&
    0&
    \cos\left(t_{\mathrm{go}}\right)\\
    \end{bmatrix}
\end{equation}
and the normalized matrices $\Bar{\Tilde{\bfB}},\, \Bar{\bfchalB},$ and $\Bar{\bfS}$ are obtained using the normalized forms of \eqref{z dot 3 DG}, \eqref{P tilde M chal_B and chal_C DG}, and \eqref{eq: determinant 2}, respectively.
Since the new normalized dynamic system is autonomous and normalized, the matrix $\Bar{\bfS}$ depends only on the normalized time-to-go, so its eigenvalues can therefore be evaluated graphically as a function of $\Bar{t}_{\mathrm{go}}$ for any altitude (i.e., $n$).
\begin{figure}[hbtp]
\centering
\includegraphics[width=0.75\textwidth]{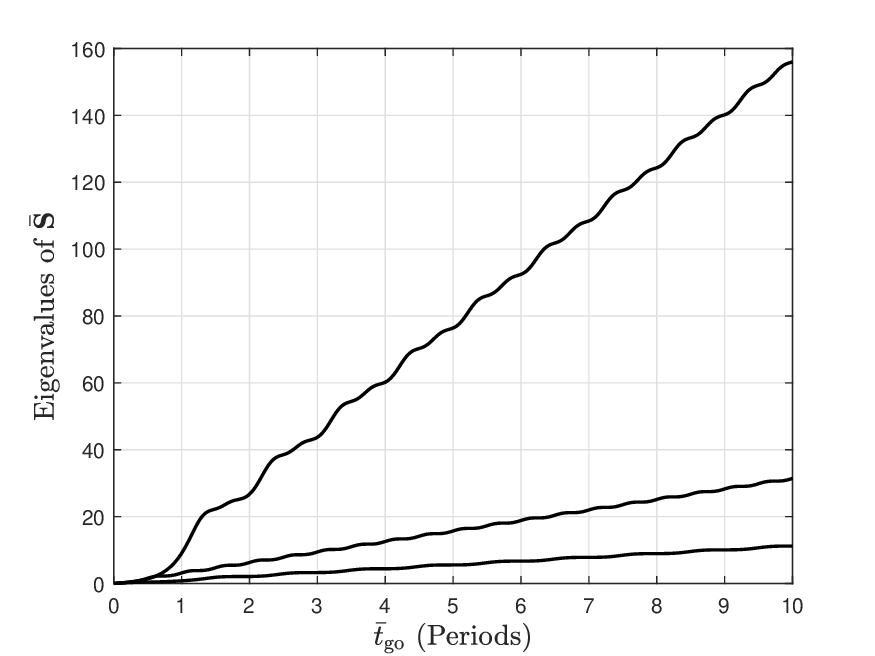}
\caption{\label{fig: S_eigenvalues}  Eigenvalues of $\Bar{\bfS}$ as function of the normalized time-to-go.}
\end{figure}

From \figref{fig: S_eigenvalues} it is evident that the eigenvalues of $\Bar{\bfS}$ are zero at $\Bar{t}_{\mathrm{go}}=0$ and monotonically increasing with $\bar{t}_{\mathrm{go}}$, and this important inference is true for any $n$.

\section*{Acknowledgments}
This work was supported by the PMRI—Peter Munk Research Institute—Technion.
\bibliography{sample}
\end{document}

%% file: command.tex
\newcommand{\bfa}{\boldsymbol{a}}
\newcommand{\bfq}{\boldsymbol{q}}

\newcommand{\bfchalB}{\boldsymbol{\mathcal{B}}}

\newcommand{\bfPhi}{\mathbf{\Phi}}

\newcommand{\bfP}{\mathbf{P}}

\newcommand{\bfA}{\mathbf{A}}
\newcommand{\bfu}{\mathbf{u}}
\newcommand{\bfr}{\mathbf{r}}
\newcommand{\bfx}{\mathbf{x}}
\newcommand{\bfv}{\mathbf{v}}
\newcommand{\bfV}{\mathbf{V}}
\newcommand{\bfI}{\mathbf{I}}
\newcommand{\bfz}{\mathbf{z}}

\newcommand{\bfQ}{\mathbf{Q}}

\newcommand{\bfB}{\mathbf{B}}

\newcommand{\bfS}{\mathbf{S}}

\newcommand{\bfLambda}{\mathbf{\Lambda}}

\newcommand{\bflambda}{\boldsymbol{\lambda}}

\newcommand{\dd}{\text{d}}

\renewcommand{\eqref}[1]{Eq.~(\ref{#1})}

\newcommand{\figref}[1]{Fig.~\ref{#1}}

\newcommand{\secref}[1]{Sec.~\ref{#1}}

\makeatletter
\renewcommand*{\p@subsection}{\thesection.}
\makeatother

\newcommand{\tref}[1]{Table~\ref{#1}}

\newcommand{\norm}[1]{\left\lVert#1\right\rVert}
